\documentclass[11pt]{article} 
\usepackage{graphicx,floatflt,amssymb,epsf,psfig,rotate,cite} 
\def\dis{\displaystyle}
\def\beq{\begin{equation}}
\def\eeq{\end{equation}}
\def\barr{\begin{array}}
\def\earr{\end{array}}

\def\gtap{\raisebox{-.4ex}{\rlap{$\sim$}} \raisebox{.4ex}{$>$}} 

\def\U{{\cal U}}
\def\O{{\cal O}}

\def\k0{K^0}
\def\k0bar{\overline{K^0}}

\begin{document} 
\begin{flushright} 
SINP/TNP/07-25
\end{flushright} 
 
\vskip 30pt 
 
\begin{center} 
{\large \bf Unraveling unparticles through violation of atomic parity
and rare beauty} \\
\vspace*{1cm} 
\renewcommand{\thefootnote}{\fnsymbol{footnote}} 
{ {\sf Gautam Bhattacharyya${}^1$}, {\sf Debajyoti Choudhury${}^{2}$} and 
{\sf Dilip Kumar Ghosh${}^{2}$} 
} \\ 
\vspace{10pt} 
{\small ${}^{1)}$ {\em Saha Institute of Nuclear Physics, 1/AF Bidhan
Nagar, Kolkata 700064, India} \\ 
${}^{2)}$ {\em Department of Physics and Astrophysics, University of
Delhi, Delhi 110 007, India}}
 
\normalsize 
\end{center} 
 
\begin{abstract} 
We put constraints on unparticle physics, specifically on the scale
$\Lambda_\U$ and the scale dimension $d_\U$ of unparticle operators,
using (i) measurements of atomic parity violation as well as (ii)
branching ratio and CP asymmetry measurements in some rare
non-leptonic $B$ decay channels.
 
\end{abstract} 

\renewcommand{\thesection}{\Roman{section}} 
\setcounter{footnote}{0} 
\renewcommand{\thefootnote}{\arabic{footnote}}

\section{Introduction}
The notion of `unparticles', recently introduced by
Georgi~\cite{Georgi:2007ek}, constitutes a new window of physics
beyond the Standard Model (SM), conceptually different from
supersymmetry and/or extra dimensions. The latter scenarios are all
about new particles more massive than what we have already encountered
in the SM. This new idea stems from the assumption that there exists a
scale invariant sector which couples to the SM. The origin of
unparticles can be traced to the degrees of freedom of this hidden
sector. We shall elaborate a bit on it later. For direct detection of
unparticles one has to rely on missing energy signals or on
distortions in momentum distributions. But even as propagators,
unparticles exhibit some properties characteristically different from
the usual expectations arising from new physics
\cite{Georgi:2007si,Cheung:2007ue}. As a result, interesting phenomena
arise when an unparticle mediated tree graph interferes with a SM
amplitude. In particular, a peculiar phase appears, only for time-like
unparticle propagator, which depends on the scale dimension of the
relevant unparticle operator.

In this Letter we study the novel aspects emerging from such unparticle
exchange in two different sectors: $(i)$ modification of
parity-violating transitions in atoms (APV), specifically, Cesium and
$(ii)$ interference of flavor-violating unparticle mediated tree graph
with the leading SM penguin diagrams in $B^\pm \to \pi^\pm K$ (quark
level $b \to s d \bar{d}$) and $B_d \to \phi K_S$ (quark level $b \to
s s \bar{s}$). While APV is mediated by space-like unparticle
propagator, the unparticle propagator for $B$ decays is time-like
which gives rise to a CP-even (strong) phase. We have chosen to focus
on these two $B$ decay modes as they do not receive any tree level SM
contributions, and hence provide interesting playgrounds for the
competition between one-loop SM and tree level new physics
contributions.  The above mentioned strong phase associated with the
unparticle propagator provides a potential source of large CP
violation that may arise out of the above interferences in $B$
decays. Conversely, a zeroish CP asymmetry, in addition to branching
ratio measurements, would put strong constraints on unparticle
parameters.

The main idea of unparticles is the following. Let us assume that at a
very high energy scale there exists a non-trivial scale invariant
sector with an infrared fixed point, whose fields will be called the
BZ fields (as first studied by Banks and Zaks
\cite{Banks:1981nn}). The SM sector interacts with the BZ sector by
the exchange of heavy messenger particles (of mass scale $M_\U$)
leading to non-renormalizable operators $\O_{\rm SM} \O_{\rm
BZ}/M_\U^k$.  The renormalizable couplings of the BZ fields then
induce dimensional transmutation~\cite{coleman-weinberg} leading to a
scale $\Lambda_\U$. Below this scale the BZ operators with scale
dimension $d_{\rm BZ}$ match to the so-called unparticle operator
$O_\U$ of scale dimension $d_\U$ yielding an effective operator
$\lambda \O_{\rm SM} \O_\U$ with a dimensionless coefficient $\lambda
\sim \Lambda_\U^{d_{\rm BZ} - d_\U}/M_\U^k$. Since the hypothetical
conformal BZ sector is self-interacting, $d_\U$ can be fractional as
well. Now, scale invariance in that sector implies that $O_\U$ does
not correspond to conventional particles, but rather describes
unparticles symbolizing a continuous mass spectrum\footnote{It has
been demonstrated~\cite{Stephanov} that the unparticle can be
deconstructed as the limiting case of an infinite tower of particles
of different masses with a regular mass spacing. This demonstration
implies a conceptual connection between unparticles and extra
dimension.}. A sizeable body of literature covering various aspects of
unparticle dynamics already exists
\cite{bulk-unp}.

The unparticles could, in principle, have any spin structure.
However, with the mass spectrum being continuous and extending to
zero, one would presume that any $\O_\U$ would be a singlet under the
SM gauge group so as to avoid profuse production of unparticles whose
detection would have been unavoidable.  In this Letter, we restrict
ourselves only to a discussion of {\em vector} unparticles. Their
couplings to the fermion currents be parametrized as
\beq
{\cal L} = \Lambda_\U^{1 - d_\U} \; \bar{f'} \, \gamma_\mu 
\left[a_L^{ff'} \; (1 - \gamma_5) + a_R^{ff'} \; (1 + \gamma_5) \right] \,f \;
{\cal O}_{\cal U}^\mu \ ,
    \label{Lagr}
\eeq
 where ${\cal O}_\U^\mu$ is a transverse and Hermitian operator of
 dimension $d_\U > 1$. Using scale invariance, the $\U$-propagator for
 time-like $P^2$ can be determined to be~\cite{Georgi:2007si,Cheung:2007ue}
\beq
\begin{array}{rcl}
\dis
\int\,e^{iPx}\,
\left\langle0\right|T({\cal O_U^\mu}(x)\,{\cal O_U^\nu}(0))
       \left|0\right\rangle\,d^4x
& = & \dis
\frac{i}{2} \, A_{d_\U}\,
\frac{-g^{\mu\nu}+P^\mu \, P^\nu/P^2}{\sin(d_\U \, \pi)}\,
\frac{1}{(P^2)^{2-d_\U}} \, e^{i \theta_\U} \, , 
\\[2ex] {\rm where}~~
A_{d_\U} & \equiv & \dis 
\frac{16 \, \pi^{5/2} } { (2 \, \pi)^{2 \, d_\U} }  \;
       \frac{ \Gamma({d_\U}+{1 \over
       2})}{\Gamma({d_\U}-1) \; \Gamma(2\,{d_\U})} \ , 
~~ \theta_\U = - d_\U \pi .   
\end{array}
    \label{propag}
\eeq
For space-like $P^2$, the propagator is real, i.e. there is no such
phase $\theta_\U$.  The expression of $A_{d_\U}$ highlights that
unparticle matter corresponds to a stream of $d_\U$ number of massless
particles, where the peculiarity is that $d_\U$ can even be a
fraction.  Eq.~(\ref{Lagr}), then, represents the effective
interaction Lagrangian, and, along with Eq.~(\ref{propag}), defines
the new physics beyond the SM. In the spirit of effective theories, we
shall consider the coefficients $a_{L,R}^{ff'}$ to be typically order
unity. Note that, in general, both flavor diagonal and nondiagonal
unparticle couplings to fermions may exist.  However, a crucial
observation is that flavor nonconserving vertices can lead to decay
processes such as $f' \to f + {\cal U}$ and, for such couplings, one
has to satisfy $d_\U > 2$ to avoid divergence of the decay width
arising from enhanced density of states in the low $P^2$
regime\footnote{Choudhury, Ghosh amd Mamta in \cite{bulk-unp}.}.
Clearly, for the calculation of APV we need flavor diagonal unparticle
current, while for $B^\pm \to \pi^\pm K$ and $B_d \to \phi K_S$ to be
induced at the tree level we have to employ flavor non-diagonal
unparticle vertices.

\section{Atomic Parity Violation (APV)}
Improved measurements of APV, mainly in Cesium (${}^{133}_{\;\;55} {\rm
Cs}$), have led to bounds on different incarnations of physics beyond
the SM (e.g. leptoquark, $R$-parity violating supersymmetry,
additional $Z'$, extra dimensions, etc \cite{apvbsm}). Such bounds are
comparable in size to the ones obtained from high energy collider
experiments. In this section we wish to explore how the exchange of
vector unparticles leads to parity-violating effects in Cesium in
addition to those already present within the SM.

For a given atom, the parity-violating electron-nucleus effects
largely accrue from the combination of the $Z$-boson's axial coupling
to the electron and vector couplings to the quarks within the
nucleus. It is conventionally parametrized in terms of the weak charge
of the nucleus $Q_W(Z,N)$. At the atomic energy scale ($\sim$ 1 MeV)
the quarks inside the nucleus act coherently and $Q_W$ is expressed as
the coherent sum of the neutral current charges of the $(2 \, Z + N)$
up-quarks and the $(2 \, N + Z)$ down-quarks in the nucleus under
question.  While the derivation of the effect within the SM can be found
in the literature~\cite{Bouchiat}, we briefly review it here, both for
the sake of completeness as well as to establish the framework for the
corresponding derivation for the case of the unparticle.

The SM and vector unparticle mediated electron-quark
interaction can be expressed in the current-current form as
\beq
\barr{rcl}
{\cal L}_{eeqq} & = & \dis \frac{e^2 \, e_q}{P^2}
\left[\bar e \; \gamma_\mu \; e \right] \; 
\left[\bar q \; \gamma^\mu \; q \right] 
+ \frac{g^2}{4 \, c_W^2 }
\left[\bar e \; \gamma_\mu \; (v_e + a_e \, \gamma_5) \; e \right] \; 
\left[\bar q \; \gamma^\mu \; (v_q + a_q \, \gamma_5) \; q \right] \; 
(P^2 - M_Z^2)^{-1}
\\[2ex]
& + & \dis 
 \left[ \bar e \, \gamma_\mu \, 
     ({\cal V}_e   + {\cal A}_e \, \gamma_5) \, e \,
\right] \;
\left[ \bar q \, \gamma^\mu \, 
      ({\cal V}_q   + {\cal A}_q \, \gamma_5) \, q \,
\right] \;
\frac{A_{d_\U}}{2} \; \Lambda_\U^{2 - 2  \, d_{\cal U}} 
\frac{(P^2)^{d_\U - 2}}{\sin \, (d_\U \, \pi)} \; , 
\earr
   \label{apv_4fermi}
\eeq
where 
\beq
\barr{rclcrcl}
v_f & \equiv & \dis  T_{3 f} - 2 \, s_W^2 \, e_f ,
& \qquad &
a_f & \equiv & \dis - T_{3 f} , 
\\[2ex]
{\cal V}_f & \equiv &  a_R^{ff} + a_L^{ff} , 
& & 
{\cal A}_f & \equiv &  a_R^{ff} - a_L^{ff} . 
\earr
\eeq
Notice that flavor diagonal unparticle mediation with quarks at one
end and electrons at the other involves space-like momentum transfer,
and hence the propagator does not involve any strong phase ($\theta_\U =
0$).  

The parity-violating part of the potential in the non-relativistic
limit arising purely from the $Z$-boson exchange is given by
\beq
\barr{rcl}
V^{(q)}_{\rm PV} ~({\rm SM})  & = 
                 &  \dis \frac{g^2}{4 \, M_W^2 } \; a_e \, v_q \;
        \left[2 \pi^2 M_Z^2 \dis{\frac{e^{ ( - M_Z \, r)}}{r}} \right]   \; 
		\left[ \vec \sigma_e \cdot \vec {\bf v}_e \right]
~\approx ~ \sqrt{2} \; G_F \; a_e \, v_q \; \delta^3(\vec {\bf r}) \;
		\left[ \vec \sigma_e \cdot \vec {\bf v}_e \right] , 
\earr
   \label{apv_pot_Z}
\eeq
where the approximation in the last step is well justified as 
$M_Z^{-1}$ is infinitesimal in comparison to the atomic length scale. 
Summing coherently the effects of all quarks in the nucleus, one can 
parametrize the APV effect in terms of the weak charge $Q_W$ of 
the nucleon which appears in the parity-violating part of the
potential of the whole nucleus as 
\beq
\barr{rcl}
V_{\rm PV} ~({\rm SM})  & =  
 &  \dis \frac{G_F}{2\sqrt{2}} \, Q_W^{\rm SM}\, \delta^3(\vec {\bf r}) \;
		\left[ \vec \sigma_e \cdot \vec {\bf v}_e \right] , 
\earr
   \label{apv_nupot_Z}
\eeq
and, to the leading order in electroweak 
theory\footnote{Radiative corrections to $Q_W^{\rm SM}$ have been
calculated~\cite{Bardin:2001ii}.}, reads 
\beq
Q_W^{\rm SM} = 2 \; \left[ (2 \, Z + N) \, v_u  + (Z + 2 \, N) \, v_d \right]
    = - N + (1 - 4 \, s^2_W) \, Z \, .  
    \label{Q:SM}
\eeq

We now estimate the unparticle  contribution 
to $Q_W$. A straightforward computation yields the following  
term for the parity-violating potential
\beq
\barr{rcl}
V_{\rm PV}(\U)  
& = & \dis  \frac{1 }{\pi \; \Lambda_\U^2} \, {\cal A}_e \,{\cal V}_q \;
      A_{d_\U} \; \Gamma(2 \, d_\U - 2) \; 
      \left\{
2 \pi^2 \Lambda_\U^2 \; (\Lambda_\U r)^{2 - 2 \, d_\U} \; 
		\frac{1}{r} \right\} 
\; 
      \left[\vec \sigma_e \cdot \vec {\bf v}_e \right]
\\[3ex]
& \approx & \dis  \frac{1}{\pi\; \Lambda_\U^2} \, {\cal A}_e \,{\cal V}_q \;
      A_{d_\U} \; \Gamma(2 \, d_\U - 2) \; 
      \delta^3(\vec r) 
      \left[ \vec \sigma_e \cdot \vec {\bf v}_e \right] ~. 
\earr
     \label{VPV_u}
\eeq
With the cutoff $\Lambda_\U > M_Z$, the delta function approximation
above has been done analogously to (but not {\em exactly} as) the SM
derivation in Eq.~(\ref{apv_pot_Z}).  The above derivation is valid as
long as $1 \leq d_\U \leq \frac{3}{2}$. Outside this range, the
calculation of $V_{\rm PV}(\U)$ needs the introduction of a regulator,
which may bring in additional model and scheme dependence\footnote{A
direct translation of APV limit from an effective contact interaction
scale to the unparticle parameter space, as inferred in Cheung, Keung
and Yuan in \cite{bulk-unp}, is not so straightforward, since the
implicit Fourier transformation with massless unparticles yields an
additional $\Gamma(2d_\U-2)$ factor, see Eq.~(\ref{VPV_u}).}. As
regards the lower limit, we recall that the very description of the
unparticle physics already restricts us to $d_\U > 1$.  Note that,
unlike in the case of many collider based observables, the factor
$\sin \, (d_\U \,
\pi)$ in the unparticle propagator cancels exactly in the expression
of the potential, thus eliminating any sharp behavior as $d_\U$
approaches integral values. Expressing the total (SM + $\U$)
contribution to APV in the form of Eq.~(\ref{apv_nupot_Z}) by
replacing $Q_W^{\rm SM}$ with $Q_W^{\rm tot} = Q_W^{\rm SM} + \delta
Q_W (\U)$, we have
\beq \dis \delta
Q_W(\U) = \frac{8 }{\Lambda_\U^2 \; G_F} \,
       \frac{(2\pi)^{3/2 -2 \, d_\U } \; \Gamma({d_\U}+{1 \over
       2})}{\Gamma({d_\U}) \; (2 \, d_\U - 1)} \; \; \; 
     {\cal A}_e \; 
\left[ (2 \, Z + N) \, {\cal V}_u + (2 \, N + Z) \, {\cal V}_d
      \right] ~. 
    \label{QW:unp}
\eeq
where the last factor in Eq.~(\ref{QW:unp}) is but a manifestation of
the coherent superposition (constructive or destructive as the case
may be).

As per the latest compilation in the Review of Particle Properties
\cite{pdg}, the experimental constraint on $Q_W$ of Cesium 
(${}^{133}_{\;\;55} {\rm Cs}$) and its SM
prediction\footnote{While experimental measurements as well as theoretical
analyses have been made for other atoms~\cite{pdg}, the current 
uncertainties are too large to compete in sensitivity.} are given by
\begin{eqnarray} 
Q_W ({\rm Expt}) & = & -72.62 \pm 0.46 \nonumber \\
Q_W ({\rm SM}) & = & -73.17 \pm 0.03 ,
\end{eqnarray} 
thus admitting a small room for new physics:  
\begin{eqnarray}
\delta Q_W \equiv Q_W({\rm Expt}) - Q_W({\rm SM}) = 0.55 \pm 0.46 \ .
     \label{delta_Q} 
\end{eqnarray}

\begin{figure}[htbp]
\begin{center}
\vspace*{-0.2cm}
\includegraphics[width= 10 cm, height= 8 cm]{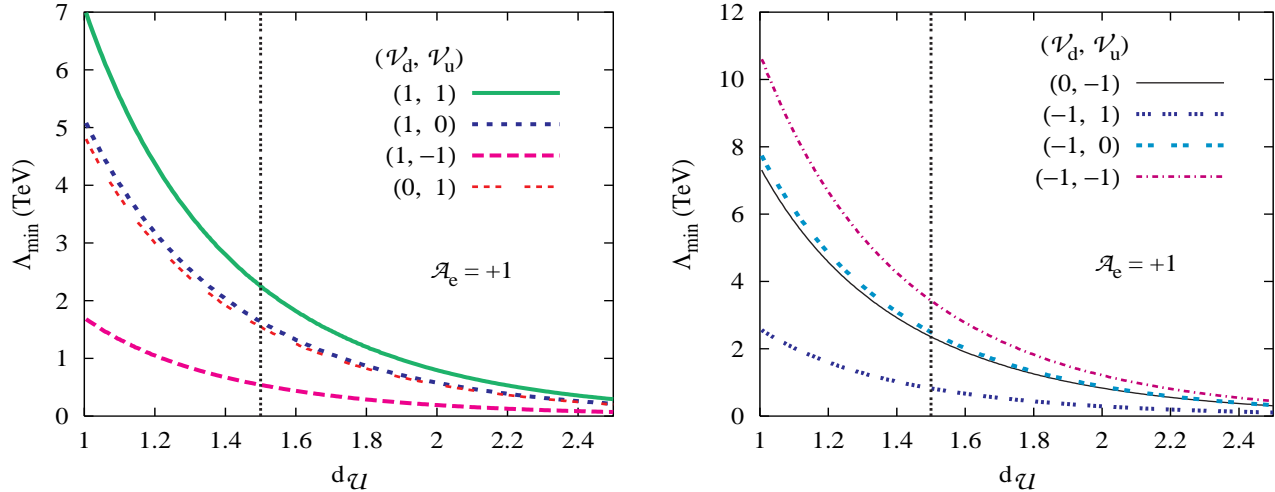}
\vspace*{-1.5cm}
\caption{\em The $3\sigma$ lower bound on $\Lambda_\U (\equiv
\Lambda_{\rm min})$ as a function 
of $d_\U$ for various combinations of the unparticle's vector
couplings to the $u$- and $d$-quark. The axial coupling to the
electron has been held to unity. For the curves in the left (right)
panel, the unparticle contribution $\delta Q_W$ is positive
(negative). The bounds to the right of the dotted vertical line
corresponds to analytic continuation of Eq.~(\ref{VPV_u}) (see text
and footnote).}
\label{fig:apv_3sig}
\end{center}
\end{figure}
Using Eqs.~(\ref{QW:unp}) and (\ref{delta_Q}), we may now derive
numerical constraints on the unparticle physics parameter space. In
Fig.~\ref{fig:apv_3sig}, we display the lower bound on $\Lambda_\U$ as
a function of $d_\U$. Since $\delta Q_W(\U)$ depends on the products
${\cal A}_e \, {\cal V}_d$ and ${\cal A}_e \, {\cal V}_u$, we may,
without any loss of generality, fix the value of one of these and we
have chosen to normalize to ${\cal A}_e = 1$.  Different combinations
of $({\cal V}_d, {\cal V}_u)$ then lead to differing constraints. For
ease of presentation, we restrict ourselves to ${\cal V}_{u,d} = 0,
\pm 1$. Thus, for our 
choice of ${\cal A}_e = 1$, a positive (negative) value of the the
combination $\left[ (2 \, Z + N) \, {\cal V}_u + (2 \, N + Z) \, {\cal
V}_d \right]$ leads to a positive (negative) $\delta Q_W$. As the
magnitude (and sign) of this combination crucially depends on those of
${\cal V}_{d, u}$, this is manifested in the relative differences in
the bounds for the various choices in Fig.~\ref{fig:apv_3sig}. Since
the only dependence (approximately exponential) on $d_\U$ is in the
pre-factor in Eq.~(\ref{QW:unp}), the shape of the curves are easily
understood. The extrapolation of curves in Fig.~\ref{fig:apv_3sig}
beyond $d_\U = 1.5$ has been achieved by analytic
continuation\footnote{By analytic continuation, we imply the existence
of a regulator that makes Eq.~(\ref{VPV_u}) to be still useable beyond
$d_\U = 3/2$ by effecting a smooth transition across the above
boundary value.}.

It should be noted that the current data prefers a small positive
$\delta Q_W$. This, obviously, can be reproduced only for certain
combinations of the couplings ${\cal A}_e$ and ${\cal V}_{u,d}$, e.g.,
those in the left panel of Fig.~\ref{fig:apv_3sig}. Finally, we
mention in passing that the unparticle contribution proportional to
${\cal V}_e\, {\cal A}_q$ probes nuclear spin, and hence the anapole
moment, which we refrain from investigating here.

\section{Rare non-leptonic $B$ decays} 
\subsection{ $B^\pm \to \pi^\pm K$} 
Thanks to the wealth of $B$-factory data, the $B^\pm \to \pi^\pm K$ decay
mode could provide important clues to unparticle parameters, in
particular, the CP-even strong phase $\theta_\U$ which is proportional
to the scale dimension $d_\U$. We consider here 
flavor violating vector unparticle current to facilitate tree level
unparticle mediated $b \to s q\bar{q}$ operator\footnote{A correlated
study of unparticle effects in $B \to \pi \pi$ and $B \to \pi K$
systems has been performed in Chen and Geng (arXiv:0706.0850 [hep-ph]) in
\cite{bulk-unp}, where flavor
diagonal vector unparticle couplings are assumed. As a result, the
leading new physics operators considered there are still penguins,
with vector unparticle replacing the SM gauge boson. Obviously, such
penguins would contribute both to $b \to s d
\bar{d}$ and $b \to s u \bar{u}$, leading to both neutral and charged
$B$ decays in all $\pi K$ modes. On the contrary, we switch on
unparticle flavor off-diagonal couplings as well which trigger just
$b \to s d \bar{d}$ interaction leading to only $B^\pm \to \pi^\pm
K$. A comparison of their results with ours is not so straightforward
as Chen and Geng (arXiv:0706.0850 [hep-ph]) in
\cite{bulk-unp} display results for $d_\U ~<~2$,
while we are constrained to take $d_\U ~>~2$ since we deal with flavor
non-diagonal unparticle couplings (see Introduction).}. We
specifically choose only those unparticle couplings which would
contribute to $b \to s d\bar{d}$ for which there is no tree diagram in
the SM. This corresponds to $B^\pm \to \pi^\pm K$, the leading SM
contribution coming from color-suppressed penguin operators. These
penguins will interfere with tree level unparticle mediated graphs
with appropriate couplings. Branching ratio and CP asymmetry
measurements in this channel would enable us to constrain $\Lambda_\U$
and $d_\U$.

The SM effective Hamiltonian for $B^+ \to \pi^+ K$ is given by 
\begin{eqnarray}
H_{\rm eff}^{\rm SM} (\pi K) & = & \frac{G_F}{\sqrt{2}} \, |V_{tb}
V^*_{ts}| \, C^{\rm SM}_{\pi K} \, f_K \, (m_B^2 - m_\pi^2) \,
F_{B\pi}
\left[1+\rho~e^{i\theta}~ e^{i\gamma}\right], 
\end{eqnarray} 
where 
\[ \dis
\rho = \left|\frac{V_{ub}
  V^*_{us}}{V_{tb}V^*_{ts}}\right| \ ,
\qquad
P \equiv \frac{m_{\k0bar}^2}{(m_s + m_d)(m_b - m_d)} \ ,
\qquad
\gamma = {\rm Arg} (V_{ub}) \;. 
\]
The SM penguin operators are captured in the combination
\[
C^{\rm SM}_{\pi K} = \frac{C_3}{3} + C_4 
   + P \left(\frac{2}{3} C_5 + 2 \, C_6 - \, \frac{C_7}{3} - C_8 \right) 
    - \, \frac{1}{2} \left( \frac{C_9}{3} + C_{10}\right) \;,  
\]
where the Wilson coefficients $C_3$--$C_{10}$ as well as the decay
constant $f_K$ and the form factor $F_{B\pi}$ may be found in
Refs.~\cite{bdk,Barberio:2007cr}.  The strong phase $\theta$ arises
from rescattering, and conservatively, is expected to be small due to
$\alpha_s$ suppression \cite{Beneke:2001ev}. We shall assume $\theta =
0$ for simplicity.

To be specific, we consider only vector unparticles, 
and switch on only the coefficients
$a_{L,R}^{bs}$ and $a_{L,R}^{bd}$.  These couplings
will induce tree level unparticle mediated $b \to s d\bar{d}$.
The total (SM + $\U$) effective Hamiltonian can now be written as 
\beq
H_{\rm eff}^{\rm tot} (\pi K) =  \frac{G_F}{\sqrt{2}} \, |V_{tb} V^*_{ts}| \,
C^{\rm SM}_{\pi K} \, f_K \, (m_B^2 - m_\pi^2) \, F_{B\pi} 
\left[1 + \rho~ e^{i\theta}~ e^{i\gamma} 
+ \rho_\U^{\pi K}~ e^{i\theta_\U}~ e^{i\gamma_\U}\right] \; , 
\eeq
where
\beq
\barr{rcl}
\rho_\U^{\pi K} & =  & \dis
\frac{- {\cal C}}{|V_{tb}V^*_{ts}| C^{\rm SM}_{\pi K}}  \, 
\left(\frac{m_b^2}{\Lambda_\U^2}\right)^{d_\U-1} \,
\frac{A_{d_\U}}{2 \sin(\pi d_\U)} \, 
\left(\frac{\sqrt{2}}{G_F m_b^2}\right) \;, 
\\[3ex]
{\rm and}~~~ {\cal C} & = & \dis 
\left[\left(a_L^{bd} a_L^{sd} - a_R^{bd} a_R^{sd}\right) 
+ 2 P \left(a_L^{bd} a_R^{sd} - a_R^{bd} a_L^{sd} \right)\right]
\\[2ex]
& + &  \dis N_c^{-1} \, \left[\left(a_L^{bs} a_L^{dd} - a_R^{bs}
a_R^{dd}\right) + 2 P \left(a_L^{bs} a_R^{dd} - a_R^{bs}
a_L^{dd}\right)\right] \; . 
\earr
    \label{C_in_bpik}
\eeq
Above, $\gamma_\U$ is a possible CP-odd weak phase associated with the
combination ${\cal C}$ of the unparticle couplings.  The unparticle
propagator is time-like in this case and gives rise to a CP-even
strong phase $\theta_\U = - d_\U \pi$. It is worth recalling that
the generation of a CP asymmetry requires both a weak phase difference
and a strong phase difference between the two interfering
amplitudes. Within our assumption of $\theta = 0$, the SM amplitude
contributes only to the weak phase ($\gamma$). The unparticle
amplitude, on the other hand, provides not only a strong
phase ($\theta_\U$), but also has the potential of contributing to the
weak phase ($\gamma_\U$). In this respect, unparticle physics scores
over several other forms of new physics in the sense that it not only
generates a tree level amplitude for $b \to s d \bar{d}$ (which
R-parity violation also does) but also provides a sizable strong phase
of a {\em different} origin.  The expression for CP asymmetry 
($\equiv [{\rm Br}~(B^+ \to f) - {\rm Br}~(B^- \to \bar{f})]/[{\rm Br}
~(B^+ \to f) + {\rm Br}~(B^- \to \bar{f})]$, with $f \equiv \pi^+ K$)
now reads
($\rho_\U \equiv \rho_\U^{\pi K}$)
\beq
\barr{rcl}
A_{\rm CP}^{\rm dir} 
& = & \dis 
\frac{2\rho \sin\theta \, \sin\gamma + 2\rho_\U \sin\theta_\U \, 
\sin\gamma_\U + 2\rho \rho_\U \sin(\theta - \theta_\U) \, \sin(\gamma -
\gamma_\U)}
{1 + \rho^2 + \rho_\U^2 + 2\rho \cos\theta \, \cos\gamma +
2\rho_\U \cos\theta_\U \, \cos\gamma_\U + 
2\rho \rho_\U \cos(\theta - \theta_\U) \, \cos(\gamma - \gamma_\U)}  
\\[3ex]
& \approx & \dis 
\frac{2\rho_\U \sin\theta_\U \, 
    \left[ \sin\gamma_\U - \rho \, \sin(\gamma -\gamma_\U)\right]}
{1 + \rho^2 + \rho_ \U^2 + 2\rho \, \cos\gamma +
2\rho_\U \cos\theta_\U \, \left[\cos\gamma_\U + 
\rho \, \cos(\gamma - \gamma_\U)\right]} 
\ . 
\earr
\eeq
where the approximate equality follows from the (excellent)
approximation of $\theta = 0$. It should be noted that the CP
asymmetry is sizable when (i) the magnitudes of the interfering
amplitudes are roughly of the same size ($\rho_\U \sim 1$, we checked,
just below $d_\U = 2.1$), and (ii) the weak and strong phase
differences between the interfering amplitudes are large.

\begin{figure}[htbp]
\begin{center}
\vspace*{0.3cm}
\includegraphics[width= 10 cm, height= 8 cm]{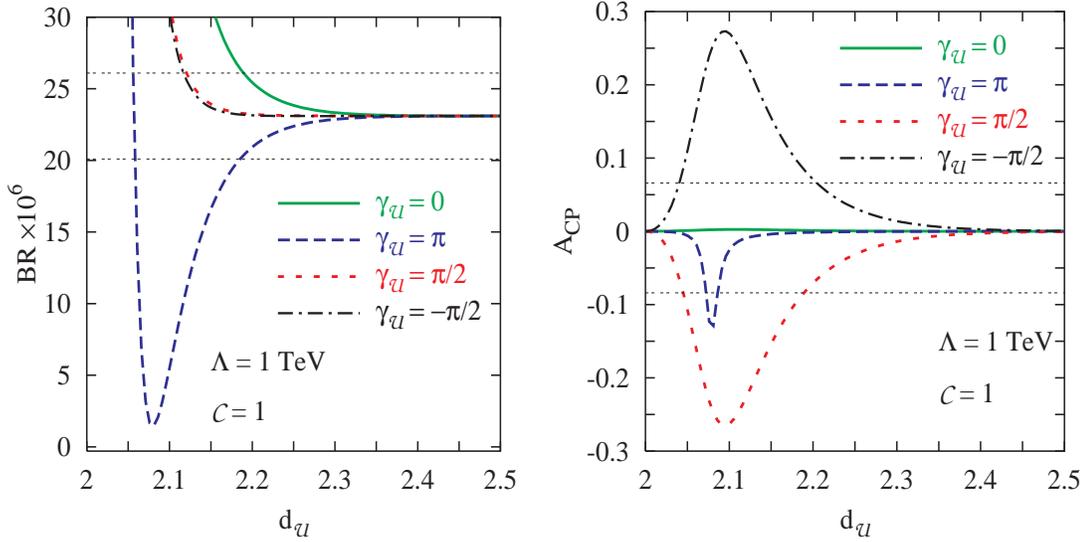}
\end{center}
\vspace*{-1.6cm}
\caption{\em The branching ratio for $B^+ \to \pi^+ K$
decays (left panel) and the direct CP asymmetry (right panel) 
as a function of $d_\U$ for $\Lambda_\U \equiv \Lambda = 1$ TeV and 
${\cal C} = 1$ (see Eq.~(\ref{C_in_bpik})). Also shown are the $3 \sigma$ 
experimental bounds~\cite{Barberio:2007cr}.}
\label{fig:b_pik}
\end{figure}

In Fig.~\ref{fig:b_pik}(left panel), we display the branching ratio ${\rm
Br}~(B^+ \to \pi^+ K)$ as a function of the scaling dimension
$d_\U$. For definiteness, we set the combination ${\cal C}$ of the
unparticle coupling constants---see Eq.~(\ref{C_in_bpik})--- to unity,
the scale of the operator to 1 TeV, and the SM 
weak phase $\gamma = 63^\circ $. The dependence on ${\cal C}$ and
$\Lambda_\U$ is rather trivial. As the left panel of
Fig.~\ref{fig:b_pik} shows, for $\gamma_\U = 0,$ and $\pm \pi/2$, the
branching fraction is a monotonic function of $d_\U$ and, for
$\Lambda_\U = 1$ TeV, becomes indistinguishable from the SM value when
$d_\U ~\gtap~ 2.3$ and $d_\U ~\gtap ~ 2.2$ respectively. While the
exact value of the SM expectations has a considerable dependence on
the hadronic matrix elements, we assume here the central value as
given in Ref.~\cite{bdk} which is quite close to the observed central
value
\cite{Barberio:2007cr}:
\beq
{\rm Br} (B^\pm \to \pi^\pm K) = (23.1 \pm 1.0) \times 10^{-6} \; . 
\eeq
For the assumed benchmark value of $\Lambda_\U$ and for $\gamma_\U =
0$, consistency with the observed branching ratio would rule out $d_\U
~<~ 2.2$ at the $3\sigma$ level.  For larger (smaller) $\Lambda_\U$,
the curve moves to the left (right). For example, $\Lambda_\U = 10$
TeV is consistent with observations down to $d_\U = 2.025$.

If we consider $\gamma_\U = \pi$ (which still leaves the unparticle
amplitude bereft of a weak phase), an interesting feature
develops. Owing to the destructive interference between the SM and the
unparticle amplitudes, the partial width now develops a minimum, and
consequently, two disjoint ranges of $d_\U$ are now consistent with
the data for a given value of $\Lambda_\U$ and the coupling
combination ${\cal C}$. As far as the partial width is concerned, the
two extreme cases $\gamma_\U = 0$ and $\gamma_\U = \pi$ constitute the
envelopes of the effect of unparticle exchange, with those for any
other choice of $\gamma_\U$ falling in between (see
Fig.~\ref{fig:b_pik}(left panel)). And, as in the case for $\gamma_\U
= 0$, all such curves move to the left as $\Lambda_\U$ is increased or
$|{\cal C}|$ is decreased.

The introduction of a weak phase in the unparticle couplings has a
much more dramatic effect in the expectations for the direct CP
asymmetry $A_{\rm CP}^{\rm dir}$. The measured value is the following
\cite{Barberio:2007cr}:
\beq
A_{\rm CP}^{\rm dir} = - 0.009 \pm 0.025 \; . 
\eeq
Again, we assume the experimental central value as the SM expectation.
As Fig.~\ref{fig:b_pik}(right panel) shows, for $\gamma_\U = 0$, the
CP-asymmetry is almost indistinguishable from the SM value. A sharp
peak in $A_{\rm CP}^{\rm dir}$ shows up for $\gamma_\U = \pi$ which
corresponds to a region of the parameter space that would lead to too
large a value for the branching ratio. A non-trivial value for
$\gamma_\U$, on the other hand, could lead to a significant
enhancement in $A_{\rm CP}^{\rm dir}$ while maintaining consistency
with the observed partial width. As expected, the cases $\gamma_\U =
\pi/2$ and $\gamma_\U = -\pi/2$ provide the envelope for $A_{CP}^{\rm
dir}$. With an increase in $\Lambda_\U$ (or, equivalently, a decrease
in ${\cal C}$), the deviation from the SM value decreases, while the
loci of the extrema move to the left. For example, $\Lambda_\U = 10$
TeV results in the maximum magnitude of $A_{CP}^{\rm dir}$ being
reduced to $-0.01 \, (+0.014)$ for $\gamma_\U = \pm \pi/2$
respectively.

\subsection{$B \to \phi K_S$} 
The quark level process for this channel is $b \to s s\bar{s}$. This
is dominated by a single amplitude in the SM and the leading
contribution comes again from penguin operators. The SM effective
Hamiltonian is given by \cite{bdk}
\beq
\barr{rcl}
H_{\rm eff}^{\rm SM} (\phi K) & = & \dis 
\frac{G_F}{\sqrt{2}} 
~|V_{tb} V^*_{ts}|~
C^{\rm SM}_{\phi K}~ f_\phi ~F_{BK} ~\lambda(m_B^2, m_\phi^2, m_K^2) ,
~~{\rm where}\\[2ex]
C^{\rm SM}_{\phi K}  &= & \dis
(C_3 + C_4) \, \left(1+ \frac{1}{N_c} \right) + C_5 + \frac{C_6}{N_c}
-  \, \frac{1}{2} \left[
C_7 + \frac{C_8}{N_c} + (C_9 + C_{10}) \left(1+ \frac{1}{N_c} \right)\right] , 
\earr
\eeq
with
$\lambda (x,y,z)  =  \sqrt{x^2 + y^2 + z^2 - 2xy -2 yz -2 zx}$. 
Once again, we consider only a vector unparticle following the structure of
Eq.~(\ref{Lagr}), but this time we switch on only the $a_{L,R}^{bs}$ and 
$a_{L,R}^{ss}$ couplings and assume them to be real. 
The total (SM + $\U$) Hamiltonian is then given by 
\beq
\barr{rcl}
H_{\rm eff}^{\rm tot} & = & H_{\rm eff}^{\rm SM} 
(1 + \rho_\U^{\phi K} ~ e^{i \theta_\U})  \; , ~~{\rm where}
\\[2ex]
\rho_\U^{\phi K} & \equiv & \dis
- \left(1+ \frac{1}{N_c} \right)
\frac{{\cal D}}{|V_{tb} \, V^*_{ts}| \; C^{\rm SM}_{\phi K}} 
\; \frac{A_{d_\U}}{2 \sin(\pi d_\U)}
\left(\frac{m_b^2}{\Lambda_\U^2}\right)^{d_\U-1} 
\left(\frac{\sqrt{2}}{G_F m_b^2}\right)  \; , 
\\[3ex]
{\rm and} ~~{\cal D} & \equiv & a_L^{bs} a_L^{ss} + a_L^{bs} a_R^{ss}
+ a_R^{bs} a_L^{ss} + a_R^{bs} a_R^{ss} \ .  
\earr
    \label{bd_phi}
\eeq

\begin{figure}[htbp]
\begin{center}
\vspace*{-0.2cm}
\includegraphics[width= 10 cm, height= 8 cm]{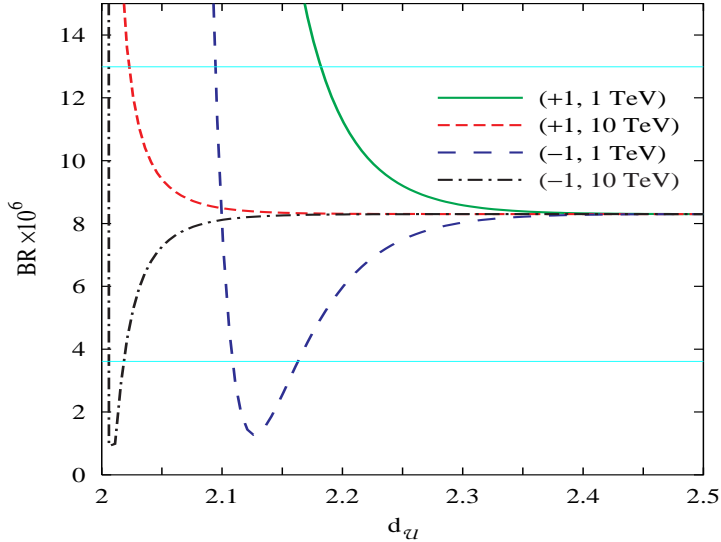}
\vspace*{-1.cm}
\end{center}
\caption{\em The branching ratio for $B^0_d \to \phi K_S$
decay as a function of $d_\U$ for different combinations of $({\cal
D}, \Lambda_\U)$ (see Eq.~(\ref{bd_phi})). Also shown are the $3 \sigma$
experimental bounds~\cite{Barberio:2007cr}.}
\label{fig:b_phiks}
\end{figure}

In Fig.~\ref{fig:b_phiks}, we display the effect of unparticle
exchange on this partial width. Once again, we use hadronic matrix
elements, as given in \cite{bdk}, and compare with the experimental
measurement \cite{Barberio:2007cr}, namely
\beq
{\rm Br}~ (B_d^0 \to \phi K_S) = \left(8.3^{+1.2}_{-1.0}\right) \times
10^{-6} \ .
\eeq
For a positive value of the coupling combination ${\cal D}$, the
branching fraction decreases monotonically with $d_\U$. As expected,
the deviation from the SM decreases with an increase in the scale
$\Lambda_\U$ (a similar behavior would be seen if the magnitude of
${\cal D}$ were to decrease). A reversal of the sign of ${\cal D}$
renders the interference between the SM and the unparticle amplitudes
destructive leading to the existence of a minimum. While we have
restricted ourselves to real-valued unparticle couplings, it is easy
to see that, for a complex unimodular ${\cal D}$, the corresponding
branching fraction would lie in between the curves for ${\cal D} = \pm
1$ acting as envelope. A CP asymmetry in this channel provides an
independent (from $B \to J/\Psi K_S$ mode) measurement of $\sin
2\beta$, but its experimental error is still too crude \cite{hfag} to
make it worthy of new physics probe.  We note in passing that by
selecting appropriate couplings and phases one can explain the current
$2 \sigma$ discrepancy between the values of $\sin2\beta$ measured
from these two modes.

\section{Conclusions} 
In this note, we have explored the effect of vector unparticles as
propagators in the atomic parity violating process as well as two rare
non-leptonic $B$ decay modes, namely, $B^\pm \to \pi^\pm K$ and $B \to
\phi K_S$. In the APV process, the virtual unparticle propagator is
space-like, while for $B$ decays it is time-like, leading to an
additional source of CP even strong phase which has a crucial impact
on the phenomenology of $B$ decays.  Moreover, while APV offers a
probe to the sensitivity of TeV scale physics through measurements at
the $\sim$ 1 MeV scale, $B$ decays could provide clues to the TeV
dynamics from measurements at a few GeV scale, so in a sense they
provide complementary hunting grounds for new physics.

With the most precise information on APV coming from the Cesium
(${}^{133}_{\;\;55} {\rm Cs}$) analysis (on account of accuracy in both
experimental measurements as well as theoretical estimates), it can be
used to obtain rather stringent conditions on the unparticle parameter
space. While the derivation of our results strictly hold only for
$d_{\U} \leq \frac{3}{2}$, they can be smoothly extended to larger
$d_{\U}$ values as well. Thus, a measurement at around 1 MeV scale is
shown to be sensitive to the unparticle scale all the way up to a few
TeV.  In particular, the discrepancy (admittedely small) between the
experimental value and the SM expectations can be explained by turning
an unparticle coupling to right-handed currents.

A flavor non-diagonal coupling of a vector unparticle with quarks
provides additional tree-level contributions to both $ B^\pm \to
\pi^\pm K$ and $B_d \to \phi K_S $ decays. For either
process, the time-like unparticle propagator gives rise to a CP even
strong phase leading to a non-trivial contribution to the
corresponding CP asymmetry.  In addition to this strong phase, a
possible weak phase $\gamma_{\U}$ may arise too if 
quark$-$unparticle couplings are complex.  Thus, the experimental data
for both the branching ratio of as well as CP asymmetry in the
$B^\pm \to \pi^\pm K$ decays may be used to impose stringent
constraints on the unparticle parameter space.  As can be expected,
such limits have a strong dependence on the value of the new weak
phase $\gamma_{\U}$. Similar results obtain for the the $B_d \to \phi
K_S$ decay process as well (although here the CP asymmetry is not a
sensitive probe as of date).

We conclude by emphasizing two points that emerge from our analyses,
and which are also shared by other authors studying other processes:
(i) besides $\Lambda_\U$, the unparticle contribution to physical
observables has a very strong dependence on $d_\U$, and (ii) flavor
off-diagonal unparticle couplings, as expected, are more strongly
constrained than the flavor diagonal ones.

\vskip 5pt
\noindent {\bf Acknowledgments:}~
GB acknowledges hospitality at CERN Theory Division and support of
the CERN Paid Associates Program. DC acknowledges support from the
Department of Science and Technology, India under project number
SR/S2/RFHEP-05/2006. DKG thanks the Theory Division, CERN and the 
HECAP Section of the AS-ICTP for hospitality while part of the work 
was completed. The authors would like to thank A. Freitas for discussions 
regarding the range of $d_U$ in Eq.~(\ref{VPV_u}).

\end{document}